# BYOD and the Mobile Enterprise

## Organisational challenges and solutions to adopt BYOD

*Andre Sobers*
Aston University
Birmingham, United Kingdom
sobersa@aston.ac.uk

*Abstract*—Bring Your Own Device, also known under the term BOYD refers to the trend in employees bringing their personal mobile devices into organisations to use as a primary device for their daily work activities. With the rapid development in computing technology in smartphones and tablet computers and innovations in mobile software and applications, mobile devices are becoming ever more powerful tools for consumers to access information. Consumers are becoming more inseparable from their personal mobile devices and development in mobile technologies within the consumer space has led to the significance of Consumerization. Enterprises everywhere want to introduce BYOD strategies to improve mobility and productivity of their employees. However making the necessary organizational changes to adopt BYOD may require a shift away from centralized systems towards more open enterprise systems and this change can present challenges to enterprises in particular over security, control, technology and policy to the traditional IT model within organisations. This paper explores some of the present challenges and solutions in relation to mobile security, technology and policy that enterprise systems within organisations can encounter. This paper also reviews real-life studies where such changes were made in organisations aiming to implement BYOD. This paper proposes a mobile enterprise model that aims to address security concerns and the challenges of technology and policy change. This paper ends with looking ahead to the future of mobile enterprise systems.

*Keywords—BYOD, MDM, Policy, Security, Technology*

## I. INTRODUCTION – THE RISE OF BYOD

Continuous improvements are being made in mobile technology, in particular with increased processing power of smartphone and tablet devices. Growth in mobile applications such as word processing, spreadsheets, presentation and other productivity tools means that mobile devices are becoming more capable of performing activities traditionally performed on a laptop or desktop computer. The popularity of owning a smartphone has risen sharply over the last few years with global sales culminating to 1.2 billion handsets in 2014 [22]. As device ownership has risen, so has the number of consumers using their devices for work. Joint research by Unisys and IDC suggests around 95% of workers have used a personal device for work [4]. This practice known as BYOD is spreading within enterprises as workers demand more BYOD support, influenced by Consumerization. While there are potential organizational benefits such as productivity, reduced costs and job satisfaction of employees, as highlighted in a 2012 Cisco global survey [23], BYOD presents new challenges to the traditional IT organisational model around areas of security, policy and technology. This paper explores these three areas with reference to relevant BYOD studies examining the challenges and changes to mobile enterprise strategies in organisations across industries from healthcare in North America [13] to manufacturing in Europe [7]. Finally this paper will look towards the future of enterprise computing strategies exploring concepts that could improve BYOD.

## II. SECURITY – PROTECTING INFORMATION ASSETS

### A. *A shift in control towards more external access to internal data*

Security remains one of the main causes of concern for IT departments who must continue to secure and protect internal data of their enterprise while ensuring employees have access to the data in order to carry out their work. According to a 2012 CompTIA study mentioned by Stagliano, DiPoalo and Coonelly [6] the survey revealed that security concerns are the greatest risk factor IT departments have when considering support for personal mobile devices. As employees are using their personal devices for work more often than devices issued by their IT department, BYOD has also shifted the balance of control from IT towards the employee.

BYOD can enable organisational data to become easily accessible to employees anywhere with an internet connection. However this transition from a closed internal 'controlled' environment towards a more open 'less-controlled' environment poses new security risks and issues on a scale not experienced by organisations before.

### B. *New horizons and new security dilemmas*

With BYOD it is possible for internal organisational data to be exposed to people outside of organisation as a result of device loss or theft. According to Symantec's 2012 Honey Stick Project, results revealed that 70% of personal devices lost were found to be directly accessing personal and corporate information by people who had found the devices [17]. Furthermore a global survey of IT professionals in 2014 run by Kaspersky Lab found that 19% of mobile device thefts led to the direct loss of corporate data [18]. Whether a mobile device is stolen or goes missing the likelihood of a missing device leaking internal data are thought to be high. Such findings emphasize the substantial risk to internal data held in organisations through the knock-on effects of personal device thefts and loss.

In addition there is the risk of internal data being leaked when an employee leaves an organisation without removing the internal data held on their device [6]. Another concern is internal data being intercepted over external wireless networks used by employees through their personal devices. Public wireless networks "are inherently less secure than a corporate network. Wireless networks are less secure because anyone who is in range of the signal of the wireless network can try to access it, whereas you would need physical access to a wired network to access it" (Stagliano, DiPoalo and Coonelly, 2012).

*C. Approaches to respond to mobile enterprise security concerns*

As a response to the security concerns Mobile Device Management (MDM) solutions are being introduced across enterprises. MDM can include the function of auto-wiping data on mobile devices as a strategy to mitigate internal data leaks [6]. Auto-wiping would clear all data on the device if for instance a maximum number of unsuccessful password attempts is reached. This method would remove any internal data on the device which would make this an effective method of protecting internal data from outsiders. However the reality with a personal device is personal data such as family photos or other media content belonging to the owner would also be lost making auto-wiping impractical and likely to face resistance of adoption from device owners.

Containerization can be considered as an alternative approach where a partition of the internal memory on a device is locked down and cannot interact with personal data that is also stored on the device. Containerization creates a protective container where internal organisational data could be stored in. A Containerization strategy could however change the user experience of the device, with reduced storage space for personal media and could limit or disable the functionality of other features on the device to the point where the owner could possibly lose interest in using their device for work.

An MDM solution could be configured to reduce intrusiveness to the end user yet provide adequate security on mobile devices. MDM could check for simple mechanisms for example ensuring a PIN or fingerprint lock is set on the device. The screen would lock after a period of inactivity and the system could attempt a remote wipe of the device if it is ever stolen or lost [1].

There is the security risk that mobile applications on personal devices may cause a leak of internal data to the outside. In 2012 IBM banned its entire 400,000 employees from using Dropbox and Apple's Siri [11]. Siri works by giving users the ability to create messages and emails through voice commands which are recorded and the data is sent to its external Apple servers for processing. Employees could be sending sensitive information outside of the network unbeknown to organisations. A possible solution to risks from mobile applications would be to incorporate a Mobile Application Management (MAM) system [10]. MAM enables "IT administrators to remotely install, update, remove, audit and monitor enterprise related applications on mobile devices" (Eslahi, Tahir, Hashim and Saad, 2014). Organisations using MAM would provide the features that could hand control back to IT departments in protecting organizational data through managing access to only 'approved' applications. MAM provides the ability to deploy patches and security updates to all mobile devices simultaneously which could improve the security of applications being used across different devices interacting with internal data.

There are potential problems with MAM [15] as "placing a virtual wall around applications can keep them from communicating with one another" (Leavitt, 2013). Furthermore this type of solution can have its limitations in that it would not "provide fine-grained control that would, for example, enable employees to access corporate information in some settings but not in others". (Leavitt, 2013).

With mobile devices employees can access internal systems through connecting through Wi-Fi however most Wi-Fi especially free public networks are unprotected. According to a 2014 Ofcom survey of businesses and consumers, around 78% admitted to mainly relying on free Wi-Fi hotspots when outside [19]. With most free Wi-Fi networks being unprotected and with growing numbers of employees accessing internal networks from outside of the office this strengthens the case for using Virtual Private Networks (VPNs) [16] while using mobile data and Wi-Fi connections [9].

Each security method mentioned has its strengths and limitations in providing protection to internal data being accessed from mobile devices. From the analysis a robust security approach should include a combination of the methods and security strategies explored earlier which should significantly reduce the security risks associated with BYOD. It is acknowledge that such a solution may not be required for some enterprises based on the nature of the data in their organisation however it represents an ideal solution for most enterprises that work with large volumes of sensitive data.

III. TECHNOLOGY – HARDWARE AND SOFTWARE CHOICES

*A. The challenges of IT supporting BYOD devices*

BYOD could bring disruption into enterprises in particular IT departments as they must extend the range of support given to mobile devices and enterprise applications. This change would cause a shift in IT budgets traditionally focused on organisation owned hardware and software. Depending on the type of BYOD strategy some organisations could face an increase in support costs in order to cover a greater range of devices. Aside from cost concerns there is also practicality challenges presented to IT which must decide the hardware and software that should be supported in a BYOD strategy.

*B. Network Access Control Technology*

Network Access Control (NAC) and MDM Technologies to support enterprises accommodating BYOD have come to the forefront of the BYOD movement [9]. NAC inspects whether the user device has complied with the organisation's security policy before it can access the internal network. NAC software controls network access according to the state of the device being unusual or suspicious. The primary goal of NAC is to block network access by infected terminals to prevent external malicious codes entering the network and spreading to other terminals potentially infecting the whole network.

The benefits of NAC is that it provides integration security functions for both wired and wireless connectivity. NAC also includes the function to authenticate mobile devices. The limitation of NAC is that it only concentrates on user authentication and access control. NAC does not possess the function to detect or respond to abnormal device behaviors once the actual device is already inside the network. Another limitation of NAC is that it focuses on authentication based on registered user devices with functionality found to be insufficient in device authentication for all devices.

*C. Mobile Device Management Technology*

MDM technology aims to provide protection for a range of channels on a mobile device that are vulnerable to data leaks such as operating apps, camera apps (taking pictures of sensitive material), audio recorder (recording internal conversations) and Wi-Fi (hacking over unprotected Wi-Fi networks). MDM also offers the control of company-wide monitoring and of the user environment through centralized management controls potentially valuable for IT departments responsible for managing corporate data security.

There are however limitations with MDM. According to a 2012 assessment of MDM, organisations may have more stringent security requirements that cannot be supported by current MDM technologies for example high security organisations which require that no data is to leave the organisation's control [20]. MDM is not capable of completely preventing data leakages from cloud services which means MDM could not fulfill this high-security requirement. Some VPN systems only focus on protecting part of the network communications between a mobile device and the internal network, in other words MDM tools do not currently protect internal data well enough in VPN systems.

The MDM agent component is not sophisticated enough to respond to all known attacks as mobile Operating System (OS) manufacturers do not share with MDM vendors the code necessary to comprehensively manage mobile devices. Different mobile device and OS producers provide different levels of code access to MDM vendors. Developers face the difficulties of providing MDM support for new OS systems as quickly as the OS is released as Apple devices for instance automatically push iOS updates to iPhones and iPads. Above all else it is difficult to distribute MDM agents onto employee devices because of reluctance in users wanting such intrusive software on their devices [9], it could raise privacy protection concerns from users.

MDM and NAC both have limitations. Independently MDM provides more comprehensive security support than NAC but still has its pitfalls. A proposed solution to bridge the gap could feature a dynamic access control system based on context information. The system would check real-time statuses of connected devices during their session on the network (e.g. context information). The system would be able to detect abnormal access and behavior of connected devices [9]. This proposed solution would remove the MDM agent concerns and also counter potential data leaks from devices already connected to the network.

*D. Supporting personal devices*

Companies may limit the types of devices IT departments support. "An organization may stipulate that only devices from a particular manufacturer are allowed, either because of security risks, hardware interoperability concerns, or a requirement for particular applications that only run on particular devices" (Marshall, 2014). At the Ottawa Hospital in Ottawa, Canada only Apple iPads and iPhones are supported in the organisation's BYOD as mobile versions of the clinical and electronic health record applications used by doctors were developed for Apple iOS devices only [13]. Considering this setup it makes sense to limit BYOD to iOS devices as applications would not work on other platforms and making applications available to be used on other platforms (e.g. Android) would incur additional costs to the enterprise. With Android OS having a more 'open' architecture framework than Apple iOS it encourages more vulnerabilities as older versions of Android that are no longer supported by Google still remain in circulation. A 2011 McAfee threat report revealed a 76% rise in malware found on devices running Android OS [21].

There is no one size fits all technology as operational needs vary between organisations. A sensible approach to deciding the technology to support BYOD must balance the desire to support a wide range of devices against interoperability and security requirements of the enterprise. Affordability of certain mobile device types and the preference of employees to own particular devices must also be taken into consideration in addition to the supporting technologies already being used in the organisation (e.g. compatibility). "Companies that use BYOD must invest in each platform/OS/carrier in their BYOD portfolio therefore this not only requires multi-device support but multi-department support as well since not everyone requires the same data but they somehow also have to communicate" (Rose, 2013).

## IV. POLICY – GUIDANCE ON BYOD

*A. Policy change to support BYOD*

Organisations have the task of implementing new BYOD policies to support the technology of mobile enterprise systems and the employees who interact through their devices. BYOD introduces new challenges to creating effective BYOD policies. For example an MDM solution on a personal device will likely entail a degree of activity monitoring and in some cases access to internal data on the personal device [7]. Such activities could risk an organisation breaching privacy rights of its employees and could leave it open to legal repercussions. For example an enterprise located in the EU, operating across different EU countries with employees using personal devices both inside and outside offices to connect to the internal network and cloud services [2]. The organisation must comply with ISO 27001 regulations, data privacy governed by the General Data Protection Regulation and must comply the different sets of regulations that each EU country has.

*B. Exploring BYOD policies for enterprises*

Unisys took the approach of creating a wide-ranging BYOD policy which aims to encompass the architecture and application of their mobile enterprise system [14]. As part of the policy employees must sign an acceptable use agreement which includes allowing Unisys to install user authentication and remote wipe software on employee devices. This expansive approach could help organisations protect internal data on employee mobile devices and maintain a degree of control over their internal data. The possible limitation of this method is user adoption. Being aware that their employer has the power to wipe all of the data on their personal device at any time, are enough employees going to sign-up to such a policy?

*C. Bring Your Own vs. Company Owned*

A proposed alternative of BYOD policy is where the company owns the mobile device referred to as "Corporate Owned, Personally Enabled" (COPE) [5]. Instead of the employee using their personal device they use a similar device issued by the employer. The device would be setup with both personal and corporate spaces each with an individual user access code. COPE would enable more organisational control which benefits enterprises. However such a strategy could entail more costs to organisations through procuring devices and setting up mobile data plans. There is also the risk that company issued devices may be unpopular with employees which could lead employees to revert to using their own personal devices instead.

Alternatively BYOD policy could incorporate measures such as terminating device access to internal data and services when an employee leaves the organisation. Employees could be issued with a SIM to use exclusively for work then return the SIM once the employee leaves the organisation [5]. This policy could encourage a higher rate of adoption since employees could use their own personal devices. However organisational support for a variety of mobile devices running different OS platforms can be expensive to manage and places greater demand on IT through technical support of these different devices and OS.

In addition to this there are several other policies with the aim of offering choice to organisations implementing BYOD. These policies are referred to as Here Is Your Own Device (HYOD) and Choose Your Own Device (CYOD) [12]. HYOD shares many similarities to COPE including the benefits of central control and limitations of user adoption. CYOD policy is different in that employees select from a list of company approved devices to choose from. Giving employees a choice can increase chances of employees finding a device of their preference, encouraging adoption and limiting the number of devices supported by IT which would be cheaper than BYOD.

A proposed alternative policy that may appeal to organisations is a context-based session policy [8]. The policy includes allocating funding limits, setting security levels and special routing techniques determined through gathering context data from internal and external sources indicating device behavior. Although it is acknowledged that organisations would need to be cautious in selecting which internal and external data is collected in order to avoid conflict with the rights of employees.

## V. THE FUTURE OF ENTERPRISE COMPUTING

As innovations in technology continue to develop at its current rate, in the near future we may witness an evolution in the underlying systems that could transform the user experience of mobile device usage at work. A potential innovation could be through developing relationships between the cyber, physical, physiological, and socio-mental spaces to replicate human intelligence within enterprise computing systems ($CP^3SME$) [24]. $CP^3SME$ is described as a "multi-dimensional complex space that generates and evolves diverse spaces to contain various types of individuals interacting with, reflecting, or influencing each other through various links within space or through spaces" (Zhuge, 2011).

Characteristics of $CP^3SME$ include multi-space situation awareness in real-time, which would enhance the virtual presence of colleagues collaborating from remote sites. Coordinating spaces could utilize the physical and socio space to help employees locate (e.g. position and elevation data) colleagues or meetings inside large office spaces. Services could be delivered to employees more effectively and efficiently through mobile applications as combining semantic linking and contextual knowledge could deliver more accurate and relevant content to employees to meet their work objectives. However $CP^3SME$ could bring new security concerns to organisations as issues traditionally isolated in older enterprise systems could propagate through multiple spaces. "One of the keys to the future enterprise computing is to build a new methodology that is able to guide the design, development and applications of the cyber-physical-social systems" (Zhuge, 2012).

## VI. CONCLUSION

BYOD is becoming more frequent in organisations over recent years. Innovations in mobile technologies in the consumer space have led to a change in user behavior with IT Consumerization influencing employees to use their personal devices for work. BYOD has brought benefits to many organisations and new challenges over existing security, technology and policy to address in mobile enterprise strategies, some of which have been explored in this paper. From the examples reviewed in this paper it is evident that BYOD will continue to be a disruption for organisations to implement whilst protecting their internal data. The most effective solution for BYOD requires combining security, technology and policy into a comprehensive framework and striking the right balance between the three. Adopting strengths from the three would provide effective data security and compliance, policy that balances the needs of the organisation with the rights of employees and technology that supports policy. In the future enterprise systems could harness information from the physical, socio and cyber space becoming more intelligent in identifying users, providing right access to organizational information, detecting and responding to threats both from outside and within the organisation.

REFERENCES

[1] J. Pinchot and K. Paullet, "Bring Your Own Devie to Work: Benefits, Security Risks and Governance Issues" Issues in Information Systems, 16(3), 2015.

[2] V. Samaras, S. Daskapan, R. Ahmad and K S Ray "An enterprise security architecture for accessing SaaS cloud services with BYOD" In Telecommunication Networks and Applications Conference (ATNAC), IEEE, pp.129 -134, 2014.

[3] C. Rose "BYOD: An examination of bring your own device in business" Review of Business Information Systems (RBIS), pp.65-70, 2013.

[4] "Do You Know Where Your Employee's Smartphone Is?", Unisy, June 28, 2010 http://www.unisys.com/news/News%20Release/Do-You-Know-Where-Your-Employees-Smartphone-Is-New-Unisys-Sponsored-Research-Shows-IT-Organizations-Are-Playing-Catch-Up-With-Rapid-Growth-of-Consumer-Technologies-in-the-Workplace [last accessed November 11, 2015].

[5] S. Dillon, F. Stahl and G. Vossen "BYOD and Governance of the Personal Cloud" International Journal of Cloud Applications and Computing (IJCAC), 2015, pp.23-35.

[6] T. Stagliano, A. DiPoalo and P. Coonelly "Consumerization of IT", 2013.

[7] R. Absalom "International Data Privacy Legislation Review: A Guide for BYOD Policies" Ovum Consulting, IT006, 234, pp.3-5, 2012.

[8] R. Copeland and N. Crespi "Analyzing consumerization-Should enterprise business context determine session policy?" In Intelligence in Next Generation Networks (ICIN), 2012 16th International Conference, IEEE, pp. 187-193, 2012.

[9] B. E. Koh, J. Oh and C. Im "A Study on Security Threats and Dynamic Access Control Technology for BYOD, Smart-work Environment" In Proceedings of the International MultiConference of Engineers and Computer Scientists, IEEE, pp. 189-192, 2014.

[10] M. Eslahi, V. M. Naseri, H. Hashim, N. Tahir, E. H. M. Saad "BYOD: Current state and security challenges" In Computer Applications and Industrial Electronics (ISCAIE), IEEE Symposium, IEEE, pp. 189-192, April 2014.

[11] G. Eschelbeck and D. Schwartzberg "BYOD risks and rewards. How to keep employee smartphones, laptops and tablets secure", 2012.

[12] A. Ghosh, P. K. Gajar, S. Rai "Bring your own device (BYOD): Security risks and mitigating strategies" Journal of Global Research in Computer Science, pp.62-70, 2013.

[13] S. Marshall "IT consumerization: A case study of BYOD in a healthcare setting" Technology Innovation Management Review, 2014.

[14] J. Burt "BYOD trend pressures corporate networks" eweek, pp.30-31, 2011.

[15] N. Leavitt "Today's mobile security requires a new approach" pp.16-19, 2013.

[16] M. A. Harris, K. P. Pattern "Mobile device security considerations for small-and medium-sized enterprise business mobility" Information Management & Computer Security, pp. 97-114, 2014.

[17] "The Symantec Smartphone Honey Stick Project", Symantec, 2014.

[18] "Kaspersky Lab Survey Shows Employees are Slow to Report Stolen Mobile Devices", Kaspersky Lab, September 2014 http://www.kaspersky.com/about/news/virus/2014/Kaspersky-Lab-Survey-Shows-Employees-are-Slow-to-Report-Stolen-Mobile-Devices [last accessed November 11, 2015].

[19] A. Scroxton "Wi-Fi users not concerned about hotspot security" http://www.computerweekly.com/news/2240226414/Wi-Fi-users-not-concerned-about-hotspot-security [last accessed November 11, 2015].

[20] "Mobile Device Management: Capability Gaps for High-Security Use Cases" NSA, 2012.

[21] "McAfee Q2 2011 Threats Report Shows Significant Growth for Malware on Mobile Platforms" McAfee, August 2011 http://www.mcafee.com/us/about/news/2011/q3/20110823-01.aspx [last accessed November 11, 2015].

[22] "Gartner Says Smartphone Sales Surpassed One Billion Units in 2014" Gartner, March 2015 http://www.gartner.com/newsroom/id/2996817 [last accessed November 11, 2015].

[23] "Cisco Study: IT Saying Yes To BYOD" Cisco, May 2012 http://newsroom.cisco.com/press-release-content?type=webcontent&articleId=854754 [last accessed November 11, 2015].

[24] H. Zhuge "Semantic linking through spaces for cyber-physical-socio intelligence: A methodology" Artificial Intelligence, pp.988-1019, 2011.

[25] H. Zhuge, "Knowledge Grid: Toward Cyber-Physical Society" World Scientific Publishing Co. 2012 (2nd ed), 2004 (1st ed).